# A Reaction-Advection-Diffusion Model to describe Non-Uniformities in Colorimetric Sensing using Thin Porous Substrates

*Kulkarni Namratha, S. Pushpavanam\**
Department of Chemical Engineering,
Indian Institute of Technology Madras-600036
\*Corresponding author email: spush@iitm.ac.in

**Abstract**

Non-uniform product (color) distribution in colorimetric paper-based sensors affects the accuracy and reliability of measurements. The underlying mechanisms responsible for this are still unclear. The coffee ring effect explains the ring-formation at the periphery. However, ring-like patterns can also be found at intermediate radial positions in these sensors. In this work, we study the influence of mass transport and reaction dynamics within porous/paper substrates on the spatial product distribution. We consider one reactant embedded in a porous substrate, which reacts with another delivered through a sessile droplet. The process is modeled in two stages. In Stage 1, droplet imbibition creates two distinct flow domains in the substrate with moving boundaries. Stage 2 commences after complete penetration. Species-substrate interactions are addressed by including a mobility factor. The developed model is used to analyze the effects of different parameters on the product distribution for two configurations, Reagent-Embedded (RE) and Analyte-Embedded (AE). Our work demonstrates ring-like patterns can form even without evaporation effects. With decreasing analyte–reagent concentration ratio, the profile shifts inward. Thicker, more porous substrates yield greater uniformity but reduce color intensity. Immobilization of embedded species enhances uniformity in RE configuration with mobile product, and in AE configuration with immobile product. The model is validated with lead and nitrite detection experiments for RE and AE configurations respectively. It successfully captures three spatial color variations observed experimentally. This study also explains the emergence of multiple rings in these systems. The insights gained are useful for optimizing sensor design and protocols for colorimetry.

# Graphical Abstract

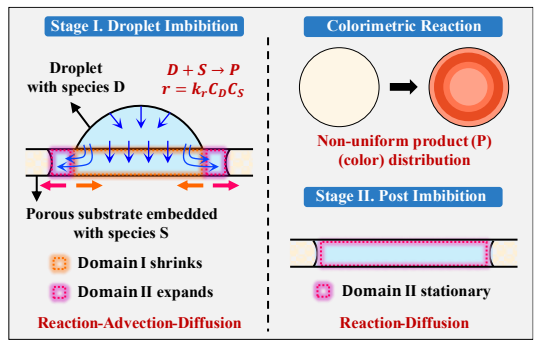

# 1. Introduction

In colorimetric sensing, the analyte (which needs to be detected and quantified) reacts with a reagent to form a colored product. The intensity of the color is used as a quantitative measure of the analyte. This method is well-suited for point-of-care (POC) applications owing to its simplicity and ability to make visual measurements with the naked eye. Microfluidic paper-based analytical devices (µPADs) or paper-based sensors have emerged as a prominent platform for colorimetric sensing in POC due to their low cost, portability, and minimal sample volumes employed. These devices have found applications in health care (detection of glucose [1] and uric acid [2]), environmental monitoring [3], and forensic analysis [4]. However, non-uniform color distribution is a major challenge in colorimetric paper-based sensors, affecting their accuracy and reproducibility [5].

In a typical colorimetric paper-based sensor, one of the reactants is uniformly embedded within a paper or a porous substrate, while the other is provided as solute in a liquid droplet. The solute in the droplet penetrates and reacts within the porous substrate to form a colored product. The distribution of the colored product depends on the hydrodynamics, mass transport of species, and reaction kinetics in the porous substrate. To mitigate non-uniform product distribution, several strategies such as the choice of paper type [6], design of reaction area [7,8], chemical modification of paper [9], and incorporation of nanomaterials [10] have been explored. These approaches optimized hydrodynamic resistance and immobilized species to enhance uniformity. Nevertheless, the underlying mechanisms responsible for the evolution of non-uniformity remain unclear.

The non-uniform product (color) distribution may appear to arise from a coffee-ring effect where solute (reactant) in the droplet accumulates at the contact line due to capillary flow. This flow is driven by non-uniform evaporation in pinned droplets on impermeable surfaces [11]. However, on permeable substrates, an additional flow arises from capillary imbibition which influences the transport of solute in the droplet. Dou and Derby (2012) demonstrated that, as the pore size increases from a few nanometers (nm) to micrometers (µm), imbibition dominates over evaporation [12]. This results in rapid absorption before significant evaporation can occur, potentially leading to more uniform solute distribution in the droplet. Despite uniform reactant imbibition from the droplet, coffee ring-like deposition can still occur in porous substrates due to the chromatographic effects, governed by solute-substrate affinity [13]. This indicates that mass transport in the porous substrate plays a major role in the non-uniform distribution of different species.

While several studies [11–14] provide insights into ring formation solely due to physical transport, the influence of chemical reactions is unexplored. Moreover, existing models do not explain the emergence of ring patterns at an intermediate radius [15]. To address this, we investigate the development of non-uniform species distributions in reactive systems on thin porous substrates using a reaction-advection-diffusion framework. In colorimetric paper-sensors, the non-uniform color distribution is produced due to the spatial distribution of the product formed in a reaction. This non-uniformity could arise due to (i) transport of the formed product within the liquid solvent present in the voids of the porous substrate, and due to (ii) spatial variation of reaction rate in the substrate. Hence, a theoretical understanding of hydrodynamics, mass transport, and reaction kinetics in thin porous substrates is necessary to understand the evolution of non-uniformities in colorimetric paper-sensors.

The hydrodynamics of droplet absorption on porous substrates has been widely studied [16–21]. On thin, dry porous substrates, the droplet motion is described by (1) spreading where the droplet base radius increases with decreasing contact angle, and (2) shrinking due to imbibition during which the base radius decreases maintaining a constant contact angle. Starov et al. (2002) experimentally and theoretically showed that the time scale of the viscous spreading is much smaller than that of imbibition, and hence the time taken for the initial spreading can be neglected [16]. Zheng (2022) modelled the hydrodynamics during the shrinking stage to describe the evolution of the penetration length in thin substrates and validated the model using experiments in literature [22]. Further, the transport of solute in thin porous substrates was studied by Wang and Darhuber (2024) in the context of inkjet printing. However, their work focuses on transport and drying after droplet imbibition and does not address solute distribution during imbibition [23]. Species transport also depends on mobility. A species may be mobile when soluble in the carrier liquid or has weak interactions with the substrate [10]. Immobility can arise from precipitation [24] or adsorption onto the substrate. These effects are considered in this study to develop a generalized framework.

Furthermore, the location of the reagent typically present in excess, governs the spatial distribution of reaction kinetics within the substrate. In conventional colorimetric sensors, the reagent is embedded in a paper substrate, and the analyte is supplied through a droplet (*Reagent-Embedded configuration*). However, this approach is ineffective for detecting trace concentrations of heavy metal ions in aqueous solutions, leading to poor measurement reliability [25,26]. Savitha et al. (2022) introduced a preconcentration-integrated sensor [27], where the analyte is embedded within a porous substrate and the reagent is supplied through a droplet (*Analyte-Embedded configuration*). In this method, the analyte is first adsorbed in batch mode followed by filtration, and the resulting residue serves as the analyte-embedded substrate. This preconcentration enhances color intensity and enables quantification at very low analyte concentrations [27]. Accordingly, we investigate both configurations in this work.

In this study, we develop a theoretical framework to explain the emergence of non-uniform product distributions in colorimetric paper-based sensors. One reactant is distributed uniformly in the paper/porous substrate, and another is supplied through a droplet which produces a colored product upon reaction. It is demonstrated that ring-like patterns can form even in the absence of evaporation effects. The analysis helps gain insights into phenomenon behind the emergence of different characteristic color distributions, ring pattern at an intermediate radius, and multiple rings. The effects of parameters such as reagent-to-analyte concentration, substrate properties, and species mobility are investigated for two reactant configurations, Reagent-Embedded and Analyte-Embedded. This study provides mechanistic insights for improving colorimetric sensing reliability.

## 2. Mathematical Model

In this work, we investigate the evolution of non-uniform product distribution in colorimetric paper-based sensors. The paper substrate is modeled as a thin porous substrate which has a uniform porosity ($\phi$), permeability ($k_p$), and thickness ($\delta$). The permeability is estimated using the Kozeny-Carman equation [28].

A reaction between two species, D and S, which form a colored product P is considered. The species S is initially distributed uniformly within the porous substrate. It reacts with the species D imbibed from a sessile droplet to form a colored product. This process is modeled in two stages: Stage 1 featuring droplet imbibition, followed by Stage 2 describing post-imbibition phase.

During Stage 1, the liquid droplet is initially assumed to be a spherical cap with an initial droplet radius $R_{d0}$ and contact angle $\theta_{a0}$. The droplet imbibes much faster than it evaporates due to the large pore size (μm) of the porous substrate [29,30]. A comparison of these timescales is presented later in this section. As the droplet imbibes into the substrate, it is assumed to maintain a constant contact angle $\theta_{a0}$ and the shape of a spherical cap, ensuring axisymmetric flow. Since the substrate is significantly thin ($\delta \ll R_{d0}$), the time for vertical saturation is negligible [22]. Hence, the variations in the vertical direction are neglected. The droplet radius $R_d$ decreases, while the wetting radius $R_p$ (the extent to which the liquid has imbibed into the porous substrate) increases with time, resulting in a moving boundary problem. In this stage, two distinct flow domains are identified in the porous medium, as depicted in **Fig. 1**(a). Domain I, $r \in [0, R_d)$ is assumed to behave like a mixed flow reactor (MFR), and in Domain II, $r \in [R_d, R_p]$ radial flow takes place. The radial flow is driven by the pressure difference across $R_d$ and $R_p$. This is caused by the capillary pressure ($p_c$) across the liquid-gas interface at $R_p$. The capillary pressure ($p_c$) is expressed using the Young-Laplace equation, $p_c = \sigma cos\theta_p/r_p$ where $\sigma$ is the surface tension, $r_p$ is the pore size, and $\theta_p$ is the angle made by the liquid with the pore walls. In this study, the wetting angle $\theta_p$ is assumed to be zero, corresponding to a completely wetting liquid.

The second stage begins after the droplet penetrates completely. Here, the pressure becomes uniform throughout the wet region (**Fig. 1**(b)) and the liquid is stationary. Since the present work is focused on sensors for point-of-care applications, we study the transport processes for a period of 5-10 min following droplet imbibition. As the amount of liquid lost due to evaporation during this time is negligible [29], the effect of evaporation on the species distribution in the porous media is neglected in this work.

The reaction between species S and D occurs within the pores and is assumed to follow second-order kinetics. One of the species is the analyte, which needs to be quantified, and the other is the reagent. Two configurations of species are considered based on the location of reactants: (i) Reagent-Embedded (RE) configuration, where the reagent is embedded in the substrate and the analyte is provided through a droplet; (ii) Analyte-Embedded (AE) configuration, where the analyte is preconcentrated in the substrate and the reagent is provided through a droplet. The analyte (A) is taken as the limiting reactant. The coffee ring effect in the droplet, which occurs due to non-uniform evaporation, is neglected as the imbibition of the droplet is very rapid compared to evaporation. We further neglect Marangoni flow in the droplet as the temperature gradient resulting from reaction is negligible. Hence, the species D is considered to be uniform in the droplet at a concentration ($C_{D_{drop}}$). In stage 1, advective, and dispersive fluxes of species are present along with reaction. Here, in domain I, species are homogeneously distributed. In stage 2, only diffusive flux is present. To account for the species-substrate interactions and solubility effects, the mobility of a species is considered. This is represented by $m_i$ which lies in the range [0,1]. It is assigned a value 1 for a fully mobile species and 0 for fully immobile species.

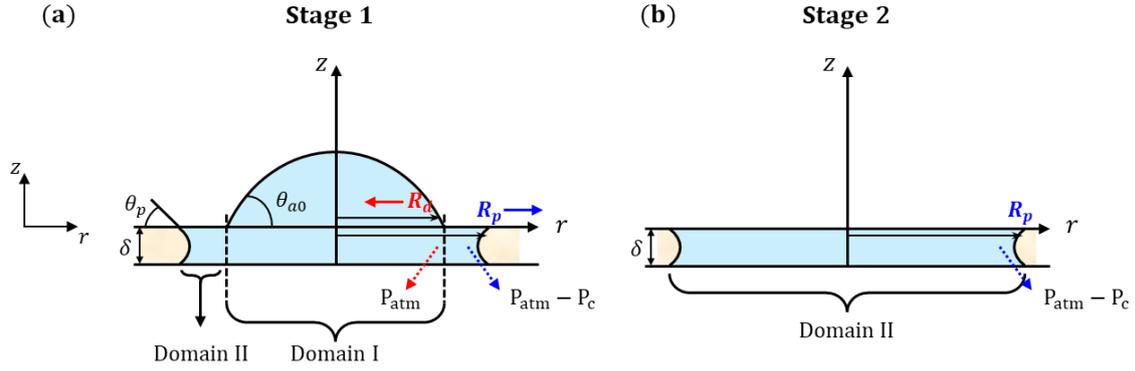

**Fig. 1** (a) Stage 1 involves droplet penetration with decreasing droplet radius, $R_d$ and increasing wetting radius, $R_p$. At $r = R_d$, the pressure in the porous substrate is continuous with the pressure in the droplet. The capillary pressure difference across the liquid interface at $R_p$ creates a pressure drop in the liquid and drives the radial flow. The wet region consists of two domains $r \in [0, R_d)$ and $r \in [R_d, R_p]$. In the first domain, species concentrations are spatially uniform. (b) Stage 2 has no liquid flow from the droplet. The pressure is constant throughout the domain $r \in [0, R_p]$.

## 2.1 Governing Equations

### 2.1.1 Stage I

*Droplet*

Stage 1 is characterized by imbibition of the sessile droplet into a thin porous substrate. The flow in the droplet is coupled with the flow in the porous substrate through mass continuity at the substrate surface [23]. The volumetric flow rate $(Q)$ at the substrate surface is the same as the rate of change of droplet volume $(V_{drop})$

$$Q(t) = -\frac{dV_{drop}}{dt} \tag{1}$$

The volume of the droplet is given by [22]

$$V_{drop} = \pi R_d^3 G(\theta_{a0}) \tag{2}$$

where $R_d$ is the droplet radius and $\theta_{a0}$ is the apparent contact angle made by the droplet with the substrate surface (z = 0). The contact angle $\theta_{a0}$ is assumed constant and $G(\theta_{a0})$ is given by

$$G(\theta_{a0}) = \frac{2 - 3\cos\theta_{a0} + \cos^3\theta_{a0}}{3\sin^3\theta_{a0}} \tag{3}$$

Thus, the volumetric flow rate $(Q)$ at the substrate surface (z = 0) is obtained as

$$Q(t) = -3\pi R_d^2 G(\theta_{a0})\frac{dR_d}{dt} \qquad (4)$$

*Porous Substrate*

The flow in the porous substrate is governed by the continuity equation

$$\vec{\nabla}.\vec{u} = 0 \qquad (5)$$

where $\vec{u}$ is the interstitial velocity. The momentum balance equation follows Darcy's law [28]

$$\vec{u} = -\frac{k_p}{\phi \mu}\vec{\nabla}p \qquad (6)$$

where $\mu$ is the viscosity of the liquid, and $k_p$, $\phi$ are the permeability and porosity of the substrate, respectively. This implies the Laplace equation is satisfied by the pressure $p$ in the porous substrate.

$$\nabla^2 p = 0 \qquad (7)$$

subject to [16,22]

$$p(r = R_p) - p(r = R_d) = -p_c \qquad (8)$$

Integrating Eq.(7) along with the above boundary condition yields the pressure gradient. Using Eq.(6), we obtain the interstitial radial velocity

$$u_r = \frac{k_p p_c}{\phi \mu}\frac{1}{r}\frac{1}{\ln\left(\frac{R_p}{R_d}\right)} \qquad (9)$$

The volumetric flow rate in the porous substrate is obtained as $Q(t) = u_r(2\pi r \delta \phi)$, i.e.,

$$Q(t) = \frac{2\pi \delta k_p p_c}{\mu}\frac{1}{\ln\left(\frac{R_p}{R_d}\right)} \qquad (10)$$

By equating Eq.(4) and Eq.(10), we obtain the evolution rate of droplet radius $R_d$ as

$$\frac{dR_d}{dt} = -\frac{2\delta k_p p_c}{3\mu G(\theta_{a0})}\frac{1}{R_d^2}\frac{1}{\ln\left(\frac{R_p}{R_d}\right)} \qquad (11)$$

The evolution rate of the wetting radius $R_p$ (the extent to which the liquid has penetrated into the porous substrate radially) is given by the radial velocity at $r = R_p$

$$\frac{dR_p}{dt} = \frac{k_p p_c}{\phi \mu}\frac{1}{R_p}\frac{1}{\ln\left(\frac{R_p}{R_d}\right)} \qquad (12)$$

In the porous substrate, as domain I $\left(r \in [0, R_d(t))\right)$ shrinks with decreasing $R_d$, the rate of change of its volume, $V_{D1}$ is given by

$$\frac{dV_{D1}}{dt} = 2\pi\delta\phi R_d \frac{dR_d}{dt} \tag{13}$$

The reaction between the species D (imbibing from droplet) and species S (embedded in the porous substrate) $D + S \rightarrow P$ is assumed elementary with a rate of

$$r_i = \nu_i k_r C_D C_S$$

Here, $\nu_i$ denotes the stoichiometric coefficient and $C_i$ denotes the concentration of species $i$ in the voids of the porous substrate. The index $i$ refers to the species $D, S,$ and P.

*Species balance equations*

The conservation of mass for each species $i \in [D, S, P]$ is governed by Eq.(14) and (15) in domains I and II respectively.

In domain I $\left(r \in [0, R_d(t))\right)$, the species concentration is assumed to be spatially uniform. On the right side of Eq. (14), the first and second terms represent the influx from the droplet and convective outflux to domain II, respectively, and the third term represents the change due to the reaction. The outflux term is multiplied by the mobility factor ($m_i$) to include convective transport only for mobile species.

$$\frac{\partial C_{i_{D1}}}{\partial t} = \frac{Q}{V_{D1}} C_{i_{drop}} - m_i \frac{Q}{V_{D1}} C_{i_{D1}} + \nu_i k_r C_{D_{D1}} C_{S_{D1}} \tag{14}$$

In domain II $\left(r \in [R_d(t), R_p(t)]\right)$, the system is governed by Eq. (15). The first term represents the net flux due to advection and diffusion, and the second term is the reaction source term.

$$\frac{\partial C_i}{\partial t} = -m_i \vec{\nabla}.\left(\vec{u} C_i - D \vec{\nabla} C_i\right) + \nu_i k_r C_D C_S \tag{15}$$

Here, the Diffusion coefficient ($D$) includes molecular diffusion $D_{m,eff}$ and dispersion. For an axisymmetric flow in the radial direction, only longitudinal dispersion $D_L$ occurs [23].

$$D = D_{m,eff} + D_L$$

At the boundaries, the influx is given by a Danckwerts boundary condition $\vec{u} C_i - D \vec{\nabla} C_i = f_i$. The flux $f_i$ is expressed at $r = R_d$ as

$$f_i|_{r=R_d} = \frac{m_i Q}{V_{D1}} C_{i_{D1}} \tag{16}$$

Accordingly, at $r = R_p$, $f_i$ is given by

$$f_i|_{r=R_p} = \begin{cases} 0 & i = D, P \\ \dfrac{1}{l_{end}} \dfrac{dR_p}{dt} C_{S0} & i = S \end{cases} \tag{17}$$

where $l_{end}$ is the length of the grid corresponding to the boundary node, $\dfrac{dR_p}{dt}$ represents the penetration rate, and $C_{S0}$ is the concentration of species S in the influx from the substrate. This flux accounts for the moles of S that have dissolved from the porous medium as the liquid penetrates.

### 2.1.2 Stage 2

Stage 2 starts after the droplet is completely imbibed. There is no driving force as the pressure becomes uniform radially. The liquid-air interface at $r = R_p$ is at rest.

$$\frac{dR_p}{dt} = 0 \tag{18}$$

The species balance equation for a species $i$ is given as

$$\frac{\partial C_i}{\partial t} = m_i D_{m,eff} \nabla^2 C_i + v_i k_r C_D C_S \tag{19}$$

In this stage, convective and dispersive mass fluxes are not present as the liquid is stationary, and only diffusive flux is present. At the boundaries, no flux condition is imposed.

### 2.2 Non-Dimensionalisation

The governing equations are made dimensionless using the following characteristic scales

$$r_{ch} = R_{d\,ch} = R_{P\,ch} = r_0, t_{ch} = \frac{\mu}{k_p} \frac{r_0^2}{p_c}, V_{ch} = \pi r_0^3 G(\theta_{a0}), u_{ch} = \frac{r_0}{t_{ch}}, Q_{ch} = \frac{V_{ch}}{t_{ch}}$$

$$C_{D\,ch} = C_{D0}, C_{S\,ch} = C_{S0}, C_{P\,ch} = C_{P0}$$

The non-dimensional variables are defined as

$$r^* = \frac{r}{r_0}, t^* = \frac{t}{t_{ch}}, R_d^* = \frac{R_d}{r_0}, R_p^* = \frac{R_p}{r_0}, C_i^* = \frac{C_i}{C_{i_{ch}}}, l_{end}^* = \frac{l_{end}}{r_0}$$

Since the substrate thickness ($\delta$) and porosity ($\phi$) can be varied experimentally, these are defined with reference values indicated by suffix 0 as

$$\alpha = \frac{\delta}{\delta_0}, \beta = \frac{\phi}{\phi_0}, \text{ and } \zeta = \frac{\dfrac{\phi^3}{(1-\phi)^2}}{\dfrac{\phi_0^3}{(1-\phi_0)^2}} \tag{20}$$

where, $\zeta$ is used to modify the permeability ($k_p$) of the porous medium.

The dimensionless parameter $\Gamma$ represents the ratio of the saturated volume of the porous substrate to the initial droplet volume at the onset of Stage 1.

$$\Gamma = \frac{2}{3} \frac{\delta_0}{r_0 G(\theta_{a0})} \tag{21}$$

Peclet number signifies the ratio of advective to diffusive mass flux, given by

$$Pe = \frac{r_0^2}{D_{m,eff} t_{ch}} \tag{22}$$

Here, the effective molecular diffusivity is written as

$$D_{m,eff} = \phi_0 D_m \tag{23}$$

and the longitudinal dispersion coefficient is expressed using the relation

$$\frac{D_L}{D_{m,eff}} = 0.5 Pe u_r^* \tag{24}$$

for the range of Pe in this study [28]. It is proportional to the Peclet number and the non-dimensional radial velocity $u_r^*$.

The non-dimensional radial velocity, flow rate, volume of domain I and its rate of change are given by

$$u_r^* = \frac{1}{\beta \phi_0 \ln\left(\frac{R_p^*}{R_d^*}\right) r^*}, \qquad Q^* = \frac{3\alpha \Gamma}{\ln\left(\frac{R_p^*}{R_d^*}\right)}$$

$$V_{D1}^* = \frac{3}{2} \alpha \Gamma \beta \phi_0 R_d^{*\,2}, \qquad \frac{dV_{D1}^*}{dt^*} = 3\alpha \Gamma \beta \phi_0 R_d^* \frac{dR_d^*}{dt^*}$$

The ratios of characteristic concentration scales of species are denoted as

$$M_A = \frac{C_{A0}}{C_{R0}}, M_P = \frac{C_{P0}}{C_{R0}} \tag{25}$$

where $C_{R0}$ and $C_{A0}$ are the characteristic scales of reagent and analyte respectively. $M_A$ indicates whether the reagent is sufficient for the analyte to react completely. When $M_A < 1$, it signifies that the reagent is in excess.

The Damkohler number is the ratio of the reaction rate to the advective transport rate of species. Since analyte is considered the limiting reactant, $Da$ is described as

$$Da = k_r C_{A0} t_{ch} \tag{26}$$

The resultant non-dimensional governing equations are:

- *Stage 1*

$$\frac{dR_d^*}{dt^*} = -\frac{\Gamma\alpha}{R_d^{*2} \ln\left(\frac{R_p^*}{R_d^*}\right)} \qquad (27)$$

$$\frac{dR_p^*}{dt^*} = \frac{1}{\beta\phi_0 R_p^* \ln\left(\frac{R_p^*}{R_d^*}\right)} \qquad (28)$$

Domain I:

$$\frac{\partial C_{i_{D1}}^*}{\partial t^*} = \frac{Q^*}{V_{D1}^*} C_{i_{drop}}^* - \frac{m_i Q^*}{V_{D1}^*} C_{i_{D1}}^* + \frac{\nu_i \eta_i}{(\alpha\beta)^2 \zeta} C_{D_{D1}}^* C_{S_{D1}}^* \qquad (29)$$

Domain II:

$$\frac{\partial C_i^*}{\partial t^*} = -m_i \vec{\nabla}^* \cdot (\vec{u}^* C_i^*) + \frac{m_i}{Pe}\frac{\beta}{\zeta} \vec{\nabla}^* \cdot \left(\left(1 + \frac{D_L}{D_{m,eff}}\right) \vec{\nabla}^* C_i^*\right) + \frac{\nu_i \eta_i}{(\alpha\beta)^2 \zeta} C_D^* C_S^* \qquad (30)$$

where, $\nabla^* = \frac{\nabla}{r_0}$ is the dimensionless operator. The boundary conditions are:

at $r^* = R_d^*$

$$f_i^*|_{r^* = R_d^*} = \frac{m_i Q^*}{V_{D1}^*} C_{i_{D1}}^* \qquad (31)$$

at $r^* = R_p^*$:

$$f_i^*|_{r^* = R_p^*} = \begin{cases} 0 & i = D, P \\ \frac{1}{l_{end}^*}\frac{dR_p^*}{dt^*} & i = S \end{cases} \qquad (32)$$

- *Stage 2*

$$\frac{\partial C_i^*}{\partial t^*} = \frac{m_i}{Pe}\frac{\beta}{\zeta} \nabla^{*2} C_i^* + \frac{\nu_i \eta_i}{(\alpha\beta)^2 \zeta} C_D^* C_S^* \qquad (33)$$

Here, $\eta_i$ is expressed as given in **Table 1** for the two configurations. Details of the calculation of $C_{R0}$, and $C_{A0}$ are provided in the supplementary information. The values of the physical parameters used in this work are given in **Table 2**. These are typical values in experiments [31]. The reactants are assumed to be dilute aqueous solutions and hence their physical properties are taken to be that of water. Details of the calculation of $G(\theta_{a0})$ are given in supplementary information.

Table 1 Values of dimensionless numbers used in this study

| Configuration | | | Da | $M_A$ | $\eta_D$ | $\eta_S$ | $\eta_P$ |
|---|---|---|---|---|---|---|---|
| Reagent – Embedded (RE) | $C_{D0} = C_{A0}$ $C_{S0} = C_{R0}$ | | $k_r t_{ch} C_{D0}$ | $\dfrac{C_{D0}}{C_{S0}}$ | $\dfrac{Da}{M_A}$ | $Da$ | $\dfrac{Da}{M_A}$ |
| Analyte-Embedded (AE) | $C_{D0} = C_{R0}$ $C_{S0} = C_{A0}$ | | $k_r t_{ch} C_{S0}$ | $\dfrac{C_{S0}}{C_{D0}}$ | $Da$ | $\dfrac{Da}{M_A}$ | $\dfrac{Da}{M_A}$ |

*2.2.1 Effect of evaporation*

During Stage 1, the sessile droplet evaporates and imbibes simultaneously. While evaporation can induce a coffee ring effect on the reagent within the droplet, imbibition influences the transport of different species in the substrate. We now analyze the relative rates of evaporation and imbibition. For this, we compare the time scales of both processes.

The imbibition time scale is obtained from Darcy's law:

$$t_{imb} = \frac{\mu r_0^2}{k_p p_c} \tag{34}$$

The time scale of evaporation is obtained by considering a quasi-steady mode [29]

$$t_{evp} = \frac{\rho V_0}{\pi r_0 D_{air}(1-H)c_v(0.27\theta^2 + 1.3)} \tag{35}$$

Using values from experiments (**Table 2**), the imbibition time scale is about 0.15 seconds, and the evaporation time is 11 minutes. The resulting ratio $\dfrac{t_{evp}}{t_{imb}} = 4.5 \times 10^3$ indicates that imbibition occurs significantly faster than droplet evaporation. Therefore, the contribution of evaporation to the species transport in the droplet can be neglected. Evaporation plays a major role in non-uniform deposition patterns on impermeable substrates.

To evaluate the significance of evaporation in Stage 2, we compare the time scales of evaporation and reaction in the system. The reaction time scale, given by $t_{ch,rxn} = 1/k_r C_{A0}$ is around 7 seconds in RE configuration and 0.7 seconds in AE configuration. The resulting ratio $\dfrac{t_{evp}}{t_{rxn}}$ is of the order of $10^2$. Therefore, we assume that the reaction is completed well before evaporation becomes relevant and neglect the evaporation effects in the system in Stage 2. This assumption is not valid for very slow reactions.

Table 2 Values of parameters used in this study

| Parameter | Notation | Value |
|---|---|---|
| Substrate thickness | $\delta_0$ | 180 μm [6] |
| Porosity | $\phi_0$ | 0.7 [31] |
| Permeability | $k_p$ | $3.1 \times 10^{-12}$ m$^2$ |
| Pore size | $r_p$ | 11 μm [6] |
| Molecular diffusivity of reagent | $D_m$ | $1.27 \times 10^{-9}$ m$^2$/s |
| Density | $\rho$ | 1000 kg/m$^3$ |
| Viscosity | $\mu$ | $10^{-3}$ Pa·s |
| Surface tension | $\sigma$ | $70 \times 10^{-3}$ N/m [32] |
| Capillary pressure | $p_c$ | $6.36 \times 10^3$ Pa |
| Initial droplet volume* | $V_0$ | 2.5 μL |
| Initial droplet radius* | $r_0$ | 1.7 mm |
| Wetting contact angle | $\theta_p$ | 0 deg |
| Reaction constant | $k_r$ | $1.5 \times 10^3$ m$^3$/(mol·s) |
| Reagent and analyte concentration * | $C_{R0}, C_{A0}$ | RE: $4.8 \times 10^{-3}$ mol/m$^3$ <br> AE: $4.8 \times 10^{-2}$ mol/m$^3$ |
| Stoichiometric coefficients | $\nu_i$ | $-1$ for $i = D, S$ <br> 1 for $i = P$ |
| Mobility of species $i$ | $m_i$ | $\in [0,1]$ <br> 1 when fully mobile <br> 0 when fully immobile |
| Relative humidity | H | 10% |
| Vapor concentration | $c_v$ | $2.32 \times 10^{-8}$ g/mm$^3$ [29] |
| Diffusion coefficient of liquid in air | $D_{air}$ | $26.1 \times 10^{-6}$ m$^2$/s [29] |

*Refer supplementary information

Table 3 Values of dimensionless numbers used in this study

| Dimensionless number | Value |
|---|---|
| $\Gamma$ | 0.795 |
| Peclet number $Pe$ | $2.2 \times 10^4$ |
| Damkohler number $Da$ | RE: 1 |
| | AE: 10 |
| Concentration ratio $M_A$ | 1 |

## 2.3 Numerical Simulation

The governing equations Eq.(27)-(33) form a set of non-linear partial differential equations (PDEs). These are solved using the Method of Lines, where the PDEs are discretized in space to obtain a set of ordinary differential equations (ODEs) in time. Domain II is discretized into 500 equidistant points, and the finite difference method is employed. Since the current system is a moving boundary problem, the grid size is updated at every time step, maintaining the total number of grids constant. The spatial derivatives are discretized using the second-order central differencing scheme at the internal points. At the boundary nodes, second-order forward and backward differencing schemes are employed. The resulting ODEs are solved using ode15s in MATLAB.

The non-dimensional initial conditions are given by

$$R_d^*(t^* = 0) = 1 \tag{36}$$

$$R_p^*(t^* = 0) = 1 + \epsilon$$

$$C_D^*(t^* = 0) = \begin{cases} 1 & \forall\, r^* \in [0,1] \\ 0 & \forall\, r^* > 1 \end{cases}$$

$$C_S^*(t^* = 0) = 1$$

$$C_P^*(t^* = 0) = 0$$

Since the substrate is thin, the time for the vertical penetration is negligibly small. Hence, at the start of stage 1, the wetting radius $R_p^*$ coincides with the droplet radius $R_d^*$. We include $\epsilon$ to avoid singularity and consider $\epsilon = 0.001$ in this work. Due to this quick vertical penetration, the species D is present in domain I at the beginning.

Stage 1 is terminated when $R_d^* \approx 0.01$. These choices for the parameters (including $\epsilon$) ensure that the overall mole balance error remains within $10^{-4}$ %. The terminal conditions of stage 1 define the initial conditions for stage 2.

## 3. Results and Discussion

We first look into the imbibition dynamics of a droplet into the thin porous substrate. This is followed by an analysis of mass transport and reaction dynamics in the porous substrate to understand the evolution of the non-uniform spatial distribution of species. The influence of reactant concentration ratio on product distribution is then investigated for both Reagent-Embedded (RE) and Analyte-Embedded (AE) configurations. Subsequently, we discuss the strategies to improve control over product distribution, including selecting an appropriate substrate and immobilizing species within the substrate. Specifically, the analysis considers two extreme mobility conditions of product (P): (1) Fully Immobile Product, (2) Fully Mobile Product. The terms "immobile" and "mobile" will hereafter refer to the fully immobile and fully mobile limits, respectively. Finally, we compare the experimental observations with model predictions for both configurations. All the results in this section are presented in dimensionless form.

### 3.1 Hydrodynamics

The droplet imbibes into the porous substrate during stage 1. Since the substrate is thin (substrate thickness ≪ initial droplet radius), the vertical penetration occurs quickly as the hydrodynamic resistance in that direction is low. This saturates the porous substrate beneath the droplet. Hence, the wetting radius $R_p^*$ (the extent to which the liquid has imbibed in the porous substrate) is equal to the initial droplet radius $R_d^*$ at $t^* = 0$. As the droplet gets imbibed, its radius $R_d^*$ decreases and the wetting radius $R_p^*$ increases due to radial flow (**Fig. 2**). As time progresses, the pressure gradient at the wetting front ($r^* = R_p^*$) decreases as shown in **Fig. 2**. This results in a reduced imbibition rate which influences the advective mass transport in the porous substrate, significantly affecting the spatial distribution of different species.

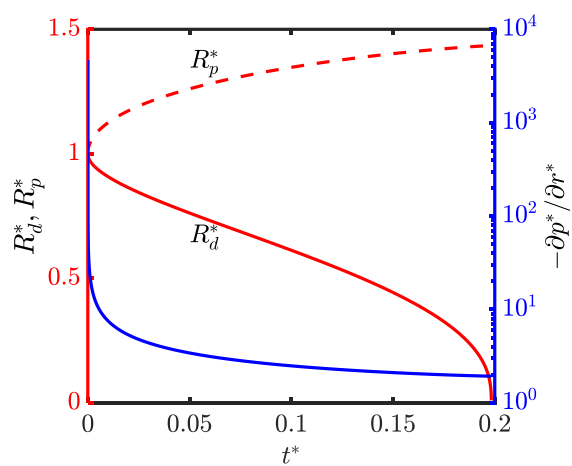

**Fig. 2** Evolution of droplet radius, $R_d^*$ and wetting radius, $R_p^*$ (Left axis); Pressure gradient ($-\partial p/\partial r$ at $r^* = R_p^*$) (Right axis) with time.

## 3.2 Product (P) Immobile

In this section, we focus on a scenario where the product (P) is immobile, while the reactants D and S, present in the droplet and substrate respectively, are mobile. The product (P) can be immobile if it is a precipitate, as observed in lead detection [27]. Additionally, strong interactions between the product and the substrate or adsorption onto the substrate can result in immobility. The reactant embedded in the substrate (S) can exhibit mobility when it has weak interactions with the substrate.

We investigate the influence of different parameters on the spatial distribution of the product. The concentration ratio of analyte to reagent ($M_A$) can be altered by varying either the reagent concentration ($C_{R0}$) or the analyte concentration ($C_{A0}$). A change in the thickness ($\delta$) and porosity ($\phi$) of the paper substrate affects the dimensionless numbers $\alpha$ and $\beta$, respectively, which influence the hydrodynamics and mass transport in the system. The effect of immobilization of the species embedded in the substrate (S) is studied by changing $m_S$. The values of dimensionless numbers (unless mentioned) are given in **Table 3**. These are obtained from the parameter values given in **Table 2**.

*3.2.1 Evolution of the spatial distribution of the product concentration (Stoichiometric ratio)*

Usually, the reagent is taken in excess to ensure that a range of analyte concentrations can be detected. The major distinction between RE and AE configurations lies in the location of the excess reagent. To begin with, we examine a scenario where the reactants are present in stoichiometric amounts, in order to investigate how spatial non-uniformity in product distribution arises. This analysis provides insights into the underlying transport phenomena common to both configurations.

**Fig. 3** illustrates the distribution of various species at different times during stage 1. The concentration is depicted over the domains I, $r^* \in [0, R_d^*(t^*))$ and II, $r^* \in [R_d^*(t^*), R_p^*(t^*)]$. The boundary of domain I, located at $r^* = R_d^*(t^*)$, is indicated by a ★ at each time instant and moves inward with time. The dashed lines represent the initial conditions. Initially, species D is present in domain I (**Fig. 3**(a)) as it quickly penetrates vertically, species S is defined till $r^* = R_p^*$, and the product is absent.

Domain I behaves as an MFR where all the species are homogenously distributed spatially; hence, there is no radial variation. There is an inflow of species D from the droplet. Also, due to advection, there is an outflow of mobile species (D and S here) from domain I to domain II. In domain I, the concentration of D ($C_D^*$) is determined by the competition between the supply rate (from droplet) and the consumption rate due to reaction and outflow. It decreases first as the reaction rate is high initially. Then it increases, since the rate of supply is more than the reaction rate. The concentration of S ($C_S^*$) decreases with time due to the reaction and outflow (**Fig. 3**(b)), while the product concentration $C_P^*$ increases (**Fig. 3**(c)).

In domain II, the mobile species (D and S here) are supplied from domain I. Additionally, as the liquid front moves forward, species S is released from the substrate into the liquid. As a result, there is a sudden increase in $C_S^*$ at $r^* = R_d^*$ (**Fig. 3**(b)). Consequently, the reaction rate increases, which is reflected by the sharp decline in $C_D^*$ at $r^* = R_d^*$ as depicted in **Fig. 3**(a). In domain II, the mobile species (D and S here) are transported radially outward by advection and dispersion. This is reflected by $C_S^*$ exceeding 1 near the periphery. Since $C_D^*$ decreases radially outward while $C_S^*$ increases, the rate of product formation has a local maximum (**Fig. 3**(c)). The position of this peak is determined by the competition between the reaction and convection. At the end of stage 1, we see a non-uniform spatial distribution in product concentration $C_P^*$.

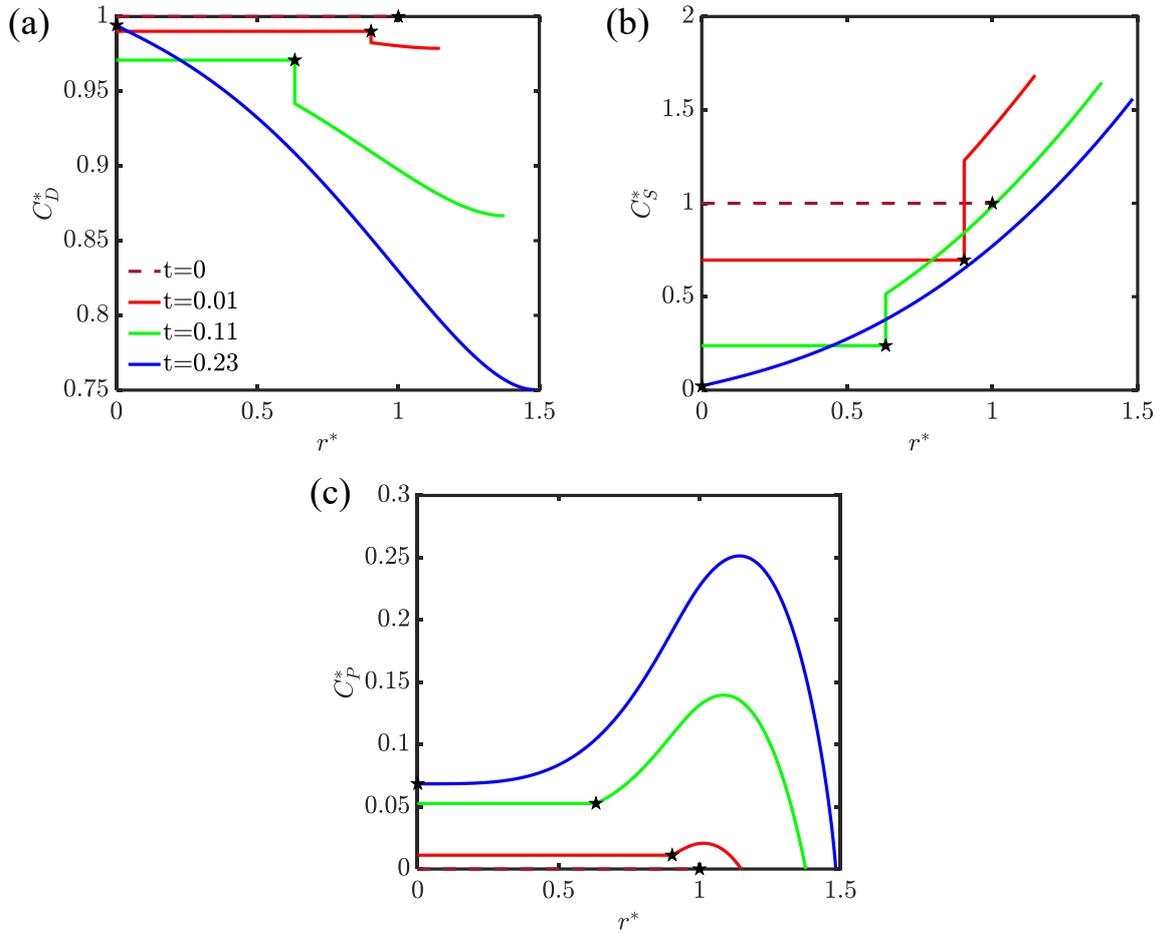

**Fig. 3** Radial distribution concentration of species (a) D, (b) S, and (c) P during stage 1 at different times (i) $t^* = 0$ (dashed line) (ii) $t^* = 0.01$ (red), (iii) $t^* = 0.1$ (green), and (iv) $t^* = 0.23$ (blue).

In stage 2 (**Fig. 4**), the liquid is stationary, and only diffusion and reaction are present. Since there is no flow, dispersion does not occur. As the reaction proceeds, the concentration of the two reactants $C_D^*$ and $C_S^*$ decreases (**Fig. 4**(a), (b)), while the product concentration $C_P^*$ increases (**Fig. 4**(c)). Since most of the species S is in the annular region by the end of stage 1 (dashed line in **Fig. 4**(b)), a significant amount of reaction occurs here. Here, a new local maximum is formed in the reaction rate based on the $C_D^*$ and $C_S^*$ profiles. Additionally, as the reaction is much faster than diffusion ($PeDa \gg 1$), the non-uniformity in product distribution persists over time.

The final product distribution has a lower concentration near the center than in the periphery. This manifests as a core region with a lower color intensity and an annular region with a higher color intensity. This captures the observed formation of ring-like dark fringes in nitrite [15] and lead detection [27]. These results are valid for both configurations as a stoichiometric reactant ratio is chosen. After some time, the reaction stops, indicating the end of stage 2. Since D is concentrated near the center and S is concentrated near the periphery, the reaction remains incomplete. Hence, excess reagent is necessary for complete consumption of the reactants.

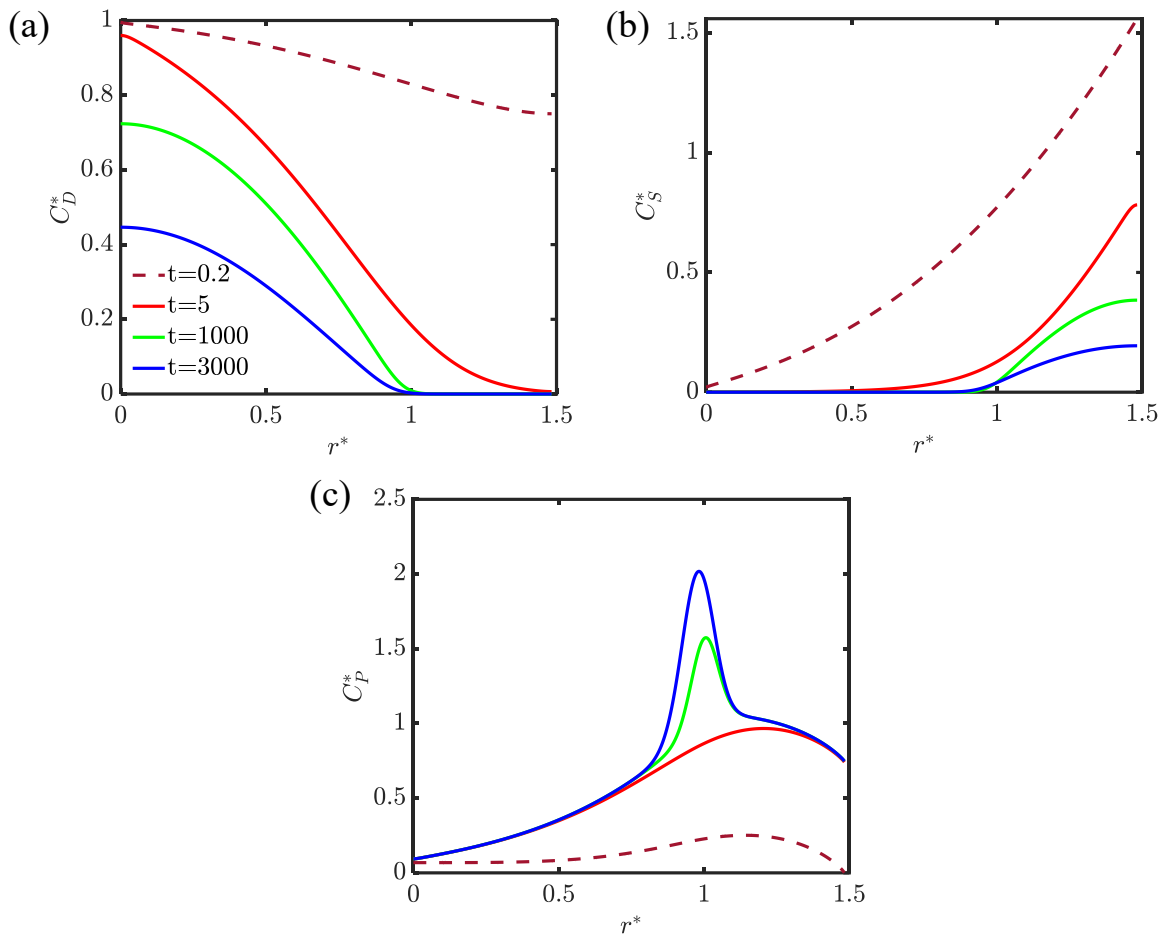

**Fig. 4** Radial distribution concentration of species (a) D, (b) S, and (c) P during stage 2 at different times (i) $t^* = 0.23$ (dashed line) (ii) $t^* = 5$ (red), (iii) $t^* = 1000$ (green), and (iv) $t^* = 3000$ (blue)

*3.2.2 Effect of concentration ratio*

The effect of the concentration ratio, $M_A = C_{A0}/C_{R0}$ on the spatial distribution of product concentration is shown in **Fig. 5** for both configurations. Here, $M_A$ is varied by changing the initial reagent concentration. The concentration of S increases radially and is higher in the annular region, while the concentration of D decreases radially (**Fig. 3** and **Fig. 4**). This results in a local maximum (peak) in the reaction rate. During stage 1, if the reaction rate is significantly higher than the transport rate, the peak shifts inward. The reaction rate is strongly affected by the concentration ratio $M_A$, influencing the location of the reaction.

In RE configuration (**Fig. 5**(a)), S is the reagent and D is the analyte. When the reagent (S) is in excess ($M_A < 1$), a considerable amount of reagent is available near the center to react with the imbibing analyte. At $M_A = 0.5$, the reagent availability is high, and hence the analyte is consumed more near the center. Consequently, the analyte transported outward is less, resulting in an inward shift of the peak (solid green line in **Fig. 5**(a)). At $M_A = 0.1$, the reagent is highly in excess, leading to increased product formation near the center during stage 1 (dashed blue line in **Fig. 5**(a)). Due to very low transport in stage 1, the analyte, which is the limiting reactant, is concentrated near the center. Hence, in stage 2, the reaction occurs primarily near the center. One more peak is formed due to a new local maximum in reaction rate (solid blue line in **Fig. 5**(a)).

Here, the concentration ratio is varied by changing the reagent concentration. A similar trend is observed when the analyte concentration is varied (see supplementary information). This analysis explains the color distributions observed by Evans et al. (2014 (a)) and Ghosh et al. (2022). In Evans et al. (2014 (a)), the reagent concentration was increased (decreased $M_A$) to minimize non-uniformity as predicted by our model. Ghosh et al. (2022) observed a ring-like color distribution in nitrite detection. This moves outward with increasing analyte concentration (increasing $M_A$) as seen in **Fig. 5**(a).

In AE configuration (**Fig. 5**(b)), S is the analyte and D is the reagent. Since the analyte (limiting reactant) is concentrated in the annular region, the reaction predominantly occurs there, leading to a ring-like product distribution. As $M_A$ decreases, the availability of the reagent increases, which increases the reaction rate. Hence, at $M_A = 0.5$ most of the reaction occurs during stage 1. Hence, a broader peak is formed, unlike the narrow peak at $M_A = 1$. When $M_A = 0.1$, the reagent is highly in excess, resulting in complete reaction during stage 1. During stage 1, the analyte located near the center is consumed more rapidly at the center rather than being transported outward. This is evident from the significant rise in product concentration near the center with decreasing $M_A$.

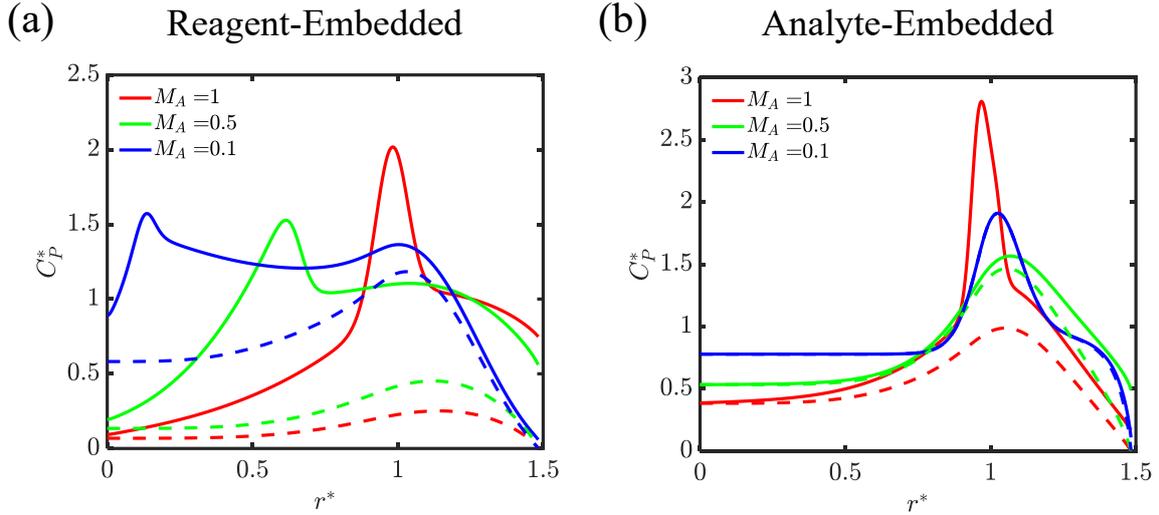

**Fig. 5** Effect of analyte-reagent concentration ratio ($M_A$) on product concentration for (a) RE, and (b) AE configurations. Here, $M_A$ is varied by varying $C_{R0}$. Dashed lines are at the end of Stage 1 and solid lines are at the end of stage 2.

*3.2.3 Effect of substrate properties*

In this subsection, we focus on the various strategies used to obtain a uniform product distribution. This is motivated by the fact that there are different papers available in the market used for paper-based colorimetric sensing. Here we investigate the effect of two key properties, substrate thickness and porosity.

*Effect of thickness*

**Fig. 6** shows the effect of thickness on product concentration distribution for both configurations. We study this for the case $M_A = 0.1$ discussed in the previous section. Here, all the concentration profiles are represented according to the concentration scale for $\alpha = 1$, where $\alpha = \delta/\delta_0$ compares the thickness $\delta$ to a reference value $\delta_0 = 180$ μm.

A change in the substrate thickness changes the volume of the porous substrate. To avoid spillage or incomplete wetting of the substrate, the droplet volume should be equal to the volume of the pores in the substrate. Accordingly, the droplet volume is taken as $V_0' = V_0 \times \alpha\beta$. Here $\beta = \phi/\phi_0$ is a dimensionless number which compares the porosity to a reference value of porosity $\beta_0 = 0.7$, hence $\beta$ lies between 0 and 10/7. With increasing thickness, the imbibition rate increases, affecting species transport.

In RE configuration, the initial concentration of reagent (S) per unit pore volume is unaltered with thickness. This is because the reagent volume used for embedding is typically equal to that of the pores in the substrate to ensure uniform distribution. When a different paper is used, the reagent volume is changed accordingly to achieve complete wetting. Hence, the reaction rate is not affected. However, higher transport rates flatten the product concentration profile with increasing thickness, as shown in **Fig. 6(a).** Here, at $\alpha = 0.5$ the peaks are steep and narrower than for $\alpha = 1$ due to low transport rates. These peaks become broader at $\alpha = 2$.

In AE configuration, the analyte (S) concentration decreases with increasing porosity. This is because the number of moles of analyte embedded is the same, while the volume of the pores in the substrate is increased. Hence, the reaction rate decreases with increasing thickness. In addition, the species transport rate increases with increasing thickness. This results in the outward shift of the peak in the product distribution profile as observed in **Fig. 6(b)**. Increasing thickness also lowers the effective product concentration, reducing color intensity. This is accompanied by a reduced color gradient in product distribution. Hence, a compromise between color intensity and color homogeneity has to be made. These findings are consistent with the experimental observations reported by Evans et al. [6].

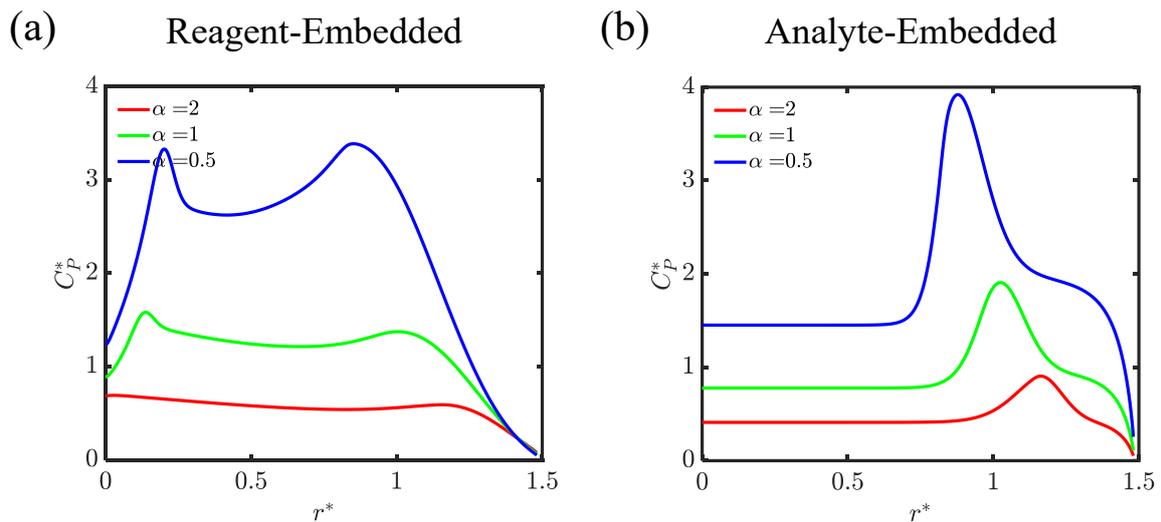

**Fig. 6** Effect of thickness on product distribution for (a) RE and (b) AE configurations

*Effect of porosity*

**Fig. 7** shows the effect of porosity on product concentration distribution for both configurations. This study is also carried out for the case $M_A = 0.1$. In **Fig. 7**, all the concentration profiles are represented according to the concentration scale for $\beta = 1$. The droplet volume is altered considering the change in volume of pores due to the change in porosity. As porosity increases, the permeability of the porous medium increases, leading to faster convection. At the same time, the reaction rate decreases due to reduced effective S concentration caused by larger substrate volume.

However, in the case of RE configuration, the reaction rate is not affected. Here at low porosity ($\beta = 5/7$), most of the product is formed near the center due to the low transport rate of the imbibing analyte (D). As a result, a radially decreasing product concentration profile is obtained. While at high porosity ($\beta = 9/7$), due to high transport rate, a more uniform product distribution is obtained.

In AE configuration, as porosity increases, the peak in the product distribution shifts outward. This is due to the increased transport rate of the analyte (S) in the substrate, along with reduced reaction rate. However, the average product concentration decreases with increasing porosity, leading to a reduced color intensity. Hence, an optimally porous substrate should be chosen to maintain balance a between color intensity and color homogeneity.

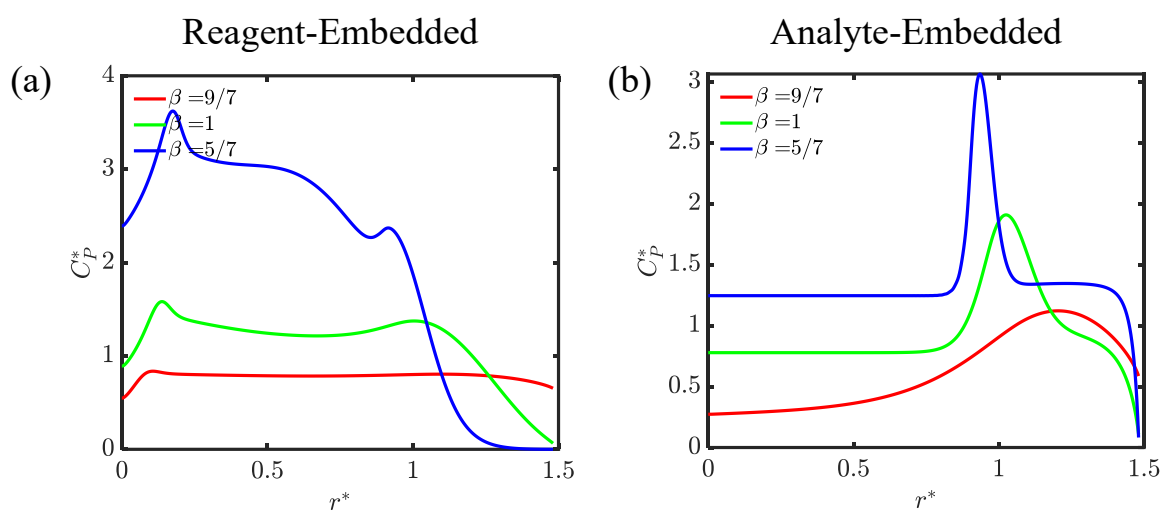

**Fig. 7** Effect of porosity on product distribution for (a) RE and (b) AE configurations

*3.2.4 Effect of immobilization of species S*

Immobilization of species S embedded in the substrate is a common strategy used to minimize non-uniformities in product distribution. This can be achieved through techniques like chemical modification of the paper and deposition of nanoparticles for enhancing the interaction between S and the substrate material. **Fig. 8** depicts the effect of mobility of S on product concentration, where the product is considered fully immobile.

In RE configuration, the behavior of product distribution depends strongly on the mobility of the reagent. When the reagent is mobile and is in a stoichiometric amount ($M_A = 1$), it is transported radially outward, leading to product formation mainly at the periphery and leaving some analyte (D) unreacted (solid red line in **Fig. 8**). Immobilizing the reagent in this case ensures complete reaction and yields a uniform product distribution (dashed red line). In practice, however, the reagent is typically used in excess. When the regent is in excess ($M_A = 0.1$) and the reagent is mobile, a product distribution with two small peaks is formed near the center due to a high reaction rate (solid blue line). If the reagent is immobilized, the spatial variation of reaction rate is governed solely by analyte transport, resulting in a radially decreasing product concentration profile (dashed blue line). These results indicate that reagent immobilization is not effective in RE configuration when the product is immobile. In AE configuration, however, immobilization remains effective, as it produces a uniform product distribution at all concentration ratios (see supplementary information).

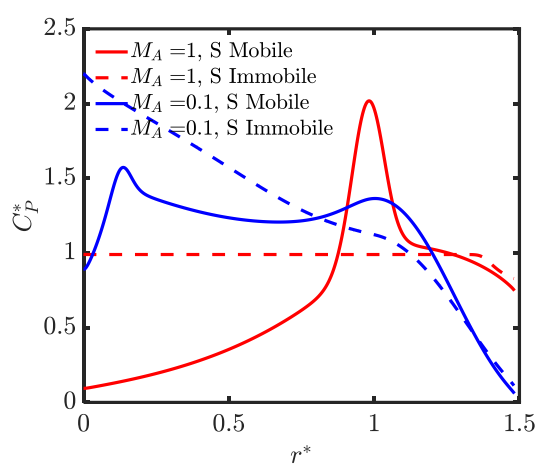

**Fig. 8** Effect of immobilization on the product distribution in RE configuration for different $M_A$.

### 3.3 Mobile Product

We now investigate the scenario where the product (P) is mobile. For this case, we study the effect of the mobility of species S at different analyte-reagent concentration ratios $M_A$.

*3.3.1 Effect of immobilization of species S (Mobile Product)*

The transport of the product (P) is dominant during stage 1 (droplet imbibition) as it gets convected and dispersed, compared to only diffusion during stage 2. If most of the product forms in stage 1, it is convected outward, leading to higher product concentration near the periphery.

In RE configuration, when both S and P are mobile, at stoichiometric conditions ($M_A = 1$), the product is formed majorly in the outer region as the reagent and analyte are transported outward while the reaction rate is low. Although, reaction rate varies spatially in the outer region, diffusion of the formed product smooths the profile (red solid curve in **Fig. 9**(a)). The product concentration is lower near the center, due to incomplete reaction, resulting in a gradient. Upon immobilization of the reagent, the reaction rate occurs throughout the area, resulting in a flatter product distribution (green dashed line). When the reagent is in excess ($M_A = 0.1$), due to the high reaction rate, most of the product is formed in domain I due to the CSTR assumption, and is transported outward during stage 1, resulting in a uniform distribution. This is observed for both S mobile and immobile conditions. Hence, when P is mobile, immobilizing the reagent helps reduce gradients across all conditions.

In AE configuration, when both the analyte (S) and product (P) are mobile, the product concentration increases radially with uniform distribution near the periphery at $M_A = 1$ (red solid line in **Fig. 9**(b)). When the reagent is in excess, most of the product forms quickly due to high reaction rate. It gets advected outside, leading to a monotonically increasing product distribution (green solid line in **Fig. 9**(b)). Immobilizing the analyte helps reduce the gradient at $M_A = 1$, but is not effective in case of excess reagent, as the product is mobile.

Hence, in RE configuration, species immobilization is effective in reducing the color gradient when the product is mobile. In the case of AE configuration, it is effective if the product is immobile.

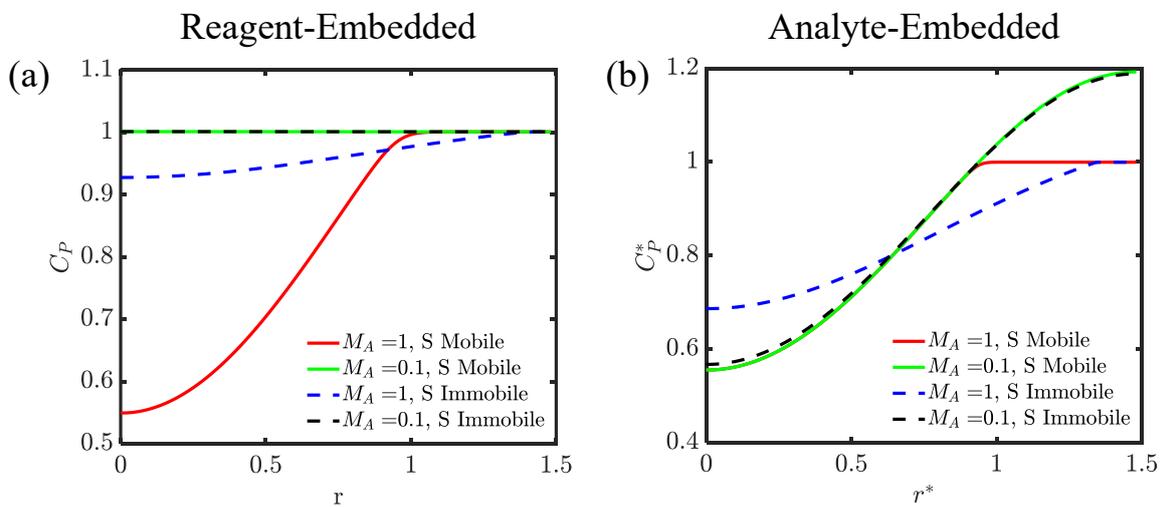

**Fig. 9** Effect of immobilization on the product distribution when product is mobile in (a) RE and (b) AE configurations for different $M_A$.

### 3.4 Experimental Methods and Measurements

In this section, we perform experimental studies to compare the observations with the predictions of the proposed model. We first investigate the spreading dynamics of a droplet on a porous substrate and validate the hydrodynamic model. Next, we perform experiments on lead ($Pb^{+2}$) detection on a kaolin-based porous substrate and nitrite ($NO_2^-$) detection on a cellulose paper representing as examples of Analyte Embedded (AE) and Reagent Embedded (RE) configurations respectively. RGB analysis is employed to quantitatively characterize the color (product) distributions. The corresponding product concentration distributions are predicted using the developed model. These are compared with the experimentally obtained color profiles to assess the model's capability in qualitatively capturing the observed trends.

*3.4.1 Hydrodynamics*

A nylon membrane filter paper (PALL, 0.45 μm pore size, 25 mm diameter, 120 μm thickness) was used as the porous substrate. and droplet imbibition process was recorded from the top using a mobile phone camera (Oppo F21 Pro). The penetration radius was measured at different time intervals. For this, the calibration was done using the diameter of the substrate. **Fig. 10**(a) shows that the model prediction shows close agreement with the measured penetration radius. Here porosity is set to 0.73 in simulation, consistent with the experimentally determined value of 0.77, validating the model. **Fig. 10**(b) further shows that the droplet is fully absorbed in ~22 seconds, after which the penetration radius remains constant. These results support the proposed two-stage imbibition framework: Stage 1 with simultaneous droplet imbibition and radial expansion of the penetration front, while Stage 2 is post complete imbibition, and here no further increase in penetration radius is observed.

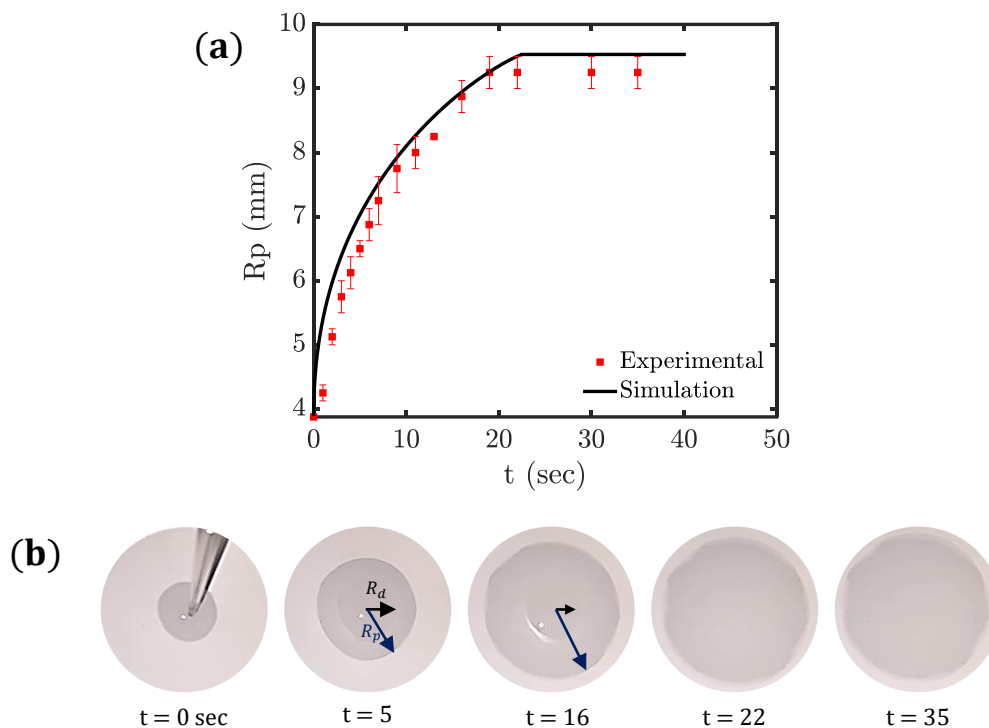

**Fig. 10** (a) Comparison of experimentally measured and predicted (simulated) penetration radius $R_p^*$ over time with error bars in the experimental measurements, (b) Droplet imbibition process with decreasing droplet radius $R_d^*$. The penetration radius $R_p^*$ increases until droplet imbibes ($t = 22$) and then stays constant.

*3.4.2 Spatial distribution of color intensity*

*I. Lead detection (AE Configuration)*

Lead ($Pb^{2+}$) in water samples is detected using sodium rhodizonate (NaRh) as a colorimetric reagent, which forms a scarlet-colored precipitate, lead rhodizonate (PbRh), upon reaction [27]. In this study, we investigate this colorimetric assay on a kaolin substrate. A stock solution of $Pb^{2+}$ was prepared and diluted to obtain solutions in the concentration range 1ppm to 100 ppm. For preconcentration, kaolin was mixed with the $Pb^{2+}$ solution and allowed to equilibrate. This was filtered through a nylon membrane to obtain a lead-enriched porous substrate. It was then dried and cooled to room temperature. The resulting substrates were 14 mm in diameter with thicknesses ranging from 110–180 µm. A solution containing the reagent NaRh was freshly prepared before each experiment, and a droplet (17 µL) was placed at the center of the substrate to produce the scarlet-colored reaction product. The process was recorded using a mobile phone positioned above the sample. More details of the experimental procedure is provided in the supplementary information.

The spatial distributions of the color (product) after reaching a steady state are compared with the model predictions for three different conditions in **Fig. 11**. The first column shows experimental images of the colored porous substrates. To quantify the observed colorimetric response, RGB of the image was analyzed. Since the green intensity ($I_{green}$) decreases as scarlet color intensity increases, a dimensionless variable is defined to represent a normalized relative color intensity.

$$\theta = \frac{255 - I_{green}}{255}$$

The variation of θ determined experimentally is shown in the second column, while in the last column, radial distributions of product concentration predicted using the developed model are shown. The three experimental conditions had varying reagent (NaRh) and analyte ($Pb^{+2}$) concentrations: (a) 100 ppm $Pb^{+2}$ with 8 mg/ml NaRh, (b) 1 ppm $Pb^{+2}$ with 1 mg/ml NaRh, (c) 1 ppm $Pb^{+2}$ with 8 mg/ml NaRh.

Parameter values used in the simulations were primarily determined from experimental measurements (Table S1), with a few chosen from experimentally reported values in the literature. The concentration scales $C_{i0}$ were calculated as explained in the supplementary information. Since PbRh, the product, is a precipitate, it is treated as immobile ($m_P = 0$). The mobility factor of analyte $m_A$ is determined to qualitatively match with the experiments. The porosity $\phi$ and pore radius $r_p$ were obtained from the literature on kaolin substrates [33]. The reaction rate constant $k_r$ was adopted from literature.

**Fig. 11** presents three distinct steady-state color patterns observed experimentally. The intensity profiles measured experimentally as well as those predicted from model are also shown. Case (a) exhibits two distinct regions, a dark core and lighter ring with decreasing intensity. The ring starts around a normalized radius ($r/r_{max}$) 0.5 as shown by the dashed line and has a width of 0.3 units. The model predicts that the ring starts around 0.3 and exists until a normalized radius of 0.6 . Beyond the ring, there is no visible product though the reagent has penetrated till the edge. The product is confined to the center, because the ratio of analyte to reagent concentration is high. So, all the reagent imbibed through the drop reacts at the central core.

Case (b) in **Fig. 11**, which occurs for a higher reagent to analyte concentration, shows three regions, a pale core, a sharp dark ring and a fading outer region. The decreasing product concentration indicates the depletion of reagent in the outer region. Experimentally, the dark ring is located between normalized radii 0.4 and 0.6 and the model predicts the ring between normalized radii 0.3 and 0.5, as shown by vertical dashed lines. We conclude that the model is able to capture the qualitative trends accurately.

In case (c) where the reagent to analyte concentration is increased further, four regions are formed. These are a pale core, a primary dark ring, a broad band of moderate-intensity and a secondary ring near the edge. The θ profile also shows two peaks. The model accurately predicts the four regions observed. The second region starts around $r/r_{max} = 0.3$, and later regions start around $r/r_{max} = 0.4$ and $0.8$ respectively.

*II. Nitrite detection (AE Configuration)*

Nitrite detection was carried out on cellulose paper using the Griess reaction. By varying the nitrite concentration from 1 to 100 mM, three distinct spatial color patterns were observed. These span center-dominated profiles to pronounced ring-like patterns. The model predictions capture the qualitative trends and the outward shift of the product distribution with increasing analyte concentration. Details of the experimental procedure, image analysis, and quantitative comparisons for this system are provided in the Supporting Information (Section S.6).

Together these results suggest that the different spatial patterns are determined from the local balance of analyte availability and reagent depletion. The proposed modeling framework is able to accurately predict these three different patterns. We emphasize that all parameters chosen for the simulations are realistic and chosen from literature.

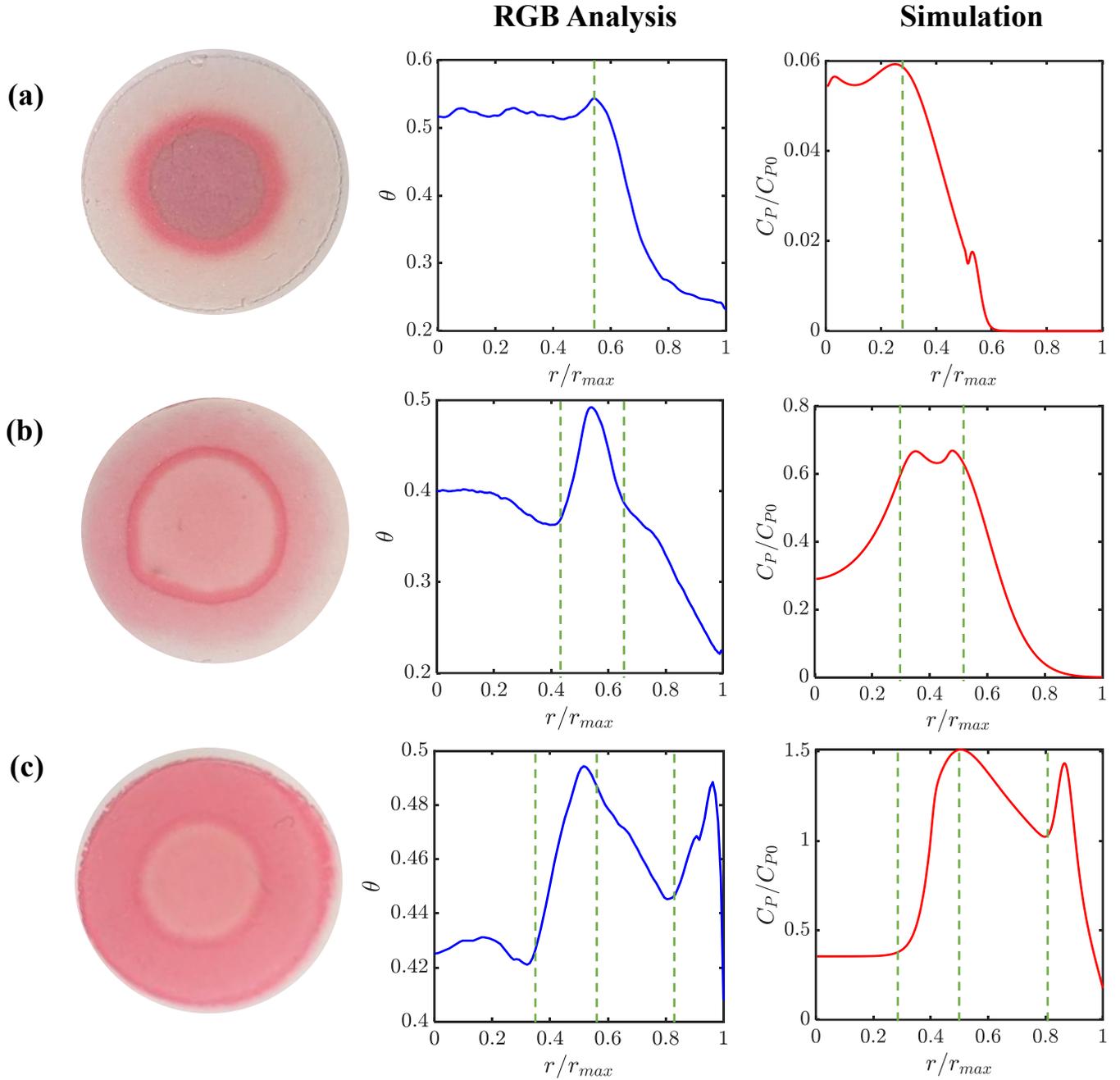

**Fig. 11** Comparison of spatial distributions of the color (product) observed experimentally in lead detection assay with model predictions for three different cases (a) 100 ppm $Pb^{+2}$ with 8 mg/ml NaRh, $C_{P0} = 1512\ mol/m^3$; (b) 1 ppm $Pb^{+2}$ with 1 mg/ml NaRh, $C_{P0} = 18\ mol/m^3$; (c) 1 ppm $Pb^{+2}$ with 8 mg/ml NaRh, $C_{P0} = 23.6\ mol/m^3$. The values used for simulations are: $\phi = 0.9$ for all, (a) $r_p = 0.05$ μm, $\delta = 180$ μm, $k_r = 0.03$ m³mol⁻¹s, $m_A = 0$, (b) $r_p = 0.03$ μm, $\delta = 150$ μm, $k_r = 0.07$ m³mol⁻¹s, $m_A = 0.7$, (c) $r_p = 0.05$ μm, $\delta = 110$ μm, $k_r = 0.03$ m³mol⁻¹s, $m_A = 0$
31

## 4. Conclusions

A theoretical model was developed to investigate the evolution of non-uniform color (product) distribution in colorimetric paper-based sensors. We model a system where one of the reactants (S) is embedded uniformly in a thin porous/paper substrate, while another reactant (D) imbibes through a sessile liquid droplet. The model is applicable for two configurations: (i) Reagent-Embedded (RE), where the reagent is embedded in the substrate (ii) Analyte-Embedded (AE), where the analyte is preconcentrated in the substrate. We incorporate the mobility factor of species to capture the effects such as washout, precipitation, and adsorption on the substrate. A species is mobile when it desorbs rapidly or is loosely attached to the substrate, and immobile if it is adsorbed on the substrate or insoluble in the liquid solvent.

We first investigated a scenario where the product P is immobile, while the species S in the substrate is mobile. We show that ring-like patterns can form even when the evaporation effects responsible for coffee-ring formation are suppressed. The transport and reaction processes within the porous substrate significantly influence the spatial distribution of the colored product. It is also shown than the ring can form at an intermediate radius, unlike the coffee ring which is formed at the periphery. The characteristics of the ring (position, width) is influenced by the competition between transport and reaction. The model developed predicts the product (color) distributions observed in lead and nitrite detection experiments. These are representative of AE and RE configurations respectively.

The parametric analysis (P immobile, S mobile) revealed that the analyte-reagent concentration ratio $M_A$ strongly influences the product distribution. For a stoichiometric reactant ratio ($M_A = 1$), the reaction remains incomplete, producing a ring-like pattern in both configurations. When the reagent is in excess ($M_A < 1$), the product concentration profile shifts inwards in AE configuration. Whereas in RE configuration, a flatter profile is obtained with decreasing $M_A$. The influence of substrate thickness and porosity on product distribution was examined as they can be experimentally varied by choosing different papers. Thicker and more porous substrates give a more uniform product distribution, but at the expense of reduced color intensity. In this study, the model also explains the emergence of multiple rings under certain conditions.

At immobile P conditions, immobilizing the species S helped reduce the product concentration gradients effectively in AE configuration, but in RE configuration it leads to radially decreasing profiles. In contrast, when the product is mobile, immobilizing S is effective in producing uniform product concentration in RE configuration but has little effect in the AE case.

This study provides valuable insights into how analyte–reagent ratio, substrate thickness, and porosity determine product (color) distributions in colorimetric sensors for varying species mobility. The model provides guidance for optimizing sensor performance and designing experimental protocols. Interestingly, it shows that non-uniform patterns on porous substrates can arise even without evaporation which is responsible for the coffee-ring phenomenon. In particular transport and reaction within the substrate govern spatial distributions.

Although the analysis focused on a second-order reaction and a circular substrate without flow barriers, the framework is readily extendable to other reaction schemes and more complex geometries, including lateral flow assays. Because species D enters the detection zone from one edge in such assays, their detection region can be viewed as analogous to half of the circular domain considered here, allowing the model to predict observed product-distribution trends.

**Conflicts of Interest**

The authors declare that they have no known competing financial interests or personal relationships that could have appeared to influence the work reported in this paper.

**Acknowledgements**

Kulkarni Namratha acknowledges the Department of Education, India for the Prime Minister Research Fellowship (PMRF).

**Supporting Information**

Calculation of $G(\theta_a)$, Determination of concentration scales $C_{R0}$ and $C_{A0}$ from experiments for both Reagent-Embedded (RE) and Analyte-Embedded (AE) configurations, Parameter values for analysis and validation studies (**Table S1**), Effect of concentration ratio $M_A$ on product distribution: Varying $C_{R0}$ (**Fig. S1**), Effect of Immobilization: AE Configuration (**Fig. S2**), Materials and Methods for lead and nitrite detection, Experimental Measurements and Comparison with Theory (RE Configuration).

# A Reaction-Advection-Diffusion Model to describe Non-Uniformities in Colorimetric Sensing using Thin Porous Substrates

*Kulkarni Namratha, S. Pushpavanam**
Department of Chemical Engineering,
Indian Institute of Technology Madras-600036
*Corresponding author email: spush@iitm.ac.in

**Supplementary Information**

### S.1 Calculation of $G(\theta_a)$

For a given initial droplet volume $V_0$, the volume balance is

$$V_0 = \pi r_0^3 G(\theta_{a0}) + \phi_0 \pi r_0^2 \delta_0 \qquad (S1)$$

where the first term denotes the volume of liquid in the spherical cap with radius $r_0$ and contact angle $\theta_{a0}$, while the second term denotes the volume of liquid in Domain I (the porous medium under the drop). Here, $\delta_0$ and $\phi_0$ are the substrate thickness and porosity respectively. $G(\theta_{a0})$ is given by

$$G(\theta_{a0}) = \frac{2 - 3\cos\theta_{a0} + \cos^3\theta_{a0}}{3\sin^3\theta_{a0}} \qquad (S2)$$

If $\theta_{a0}$ is known, this expression Eq.(S2) can be used to calculate $G(\theta_{a0})$. If it is not known, Eq.(S1) can be used to determine $\theta_{a0}$ provided $r_0$ is known.

### S.2 Determination of concentration scales $C_{R0}$ and $C_{A0}$ from experiments

*S.2.1 Reagent-Embedded (RE) Configuration*

The reagent is embedded uniformly in the paper substrate by adding a droplet of reagent and allowing the substrate to dry. Suppose the concentration of the reagent is $C_{reag}$ (in mg/ml) and the volume of the reagent used for embedding is $V_{reag}$ in ml. Then the number of moles embedded is

$$N_{R0} = \frac{C_{reag} \times V_{reag}}{M_{W_R} \times 10^3} \qquad (S3)$$

where $M_{W_R}$ is the molecular weight of reagent in g/mol. The concentration scale of reagent $C_{R0}$ is defined as the number of moles per unit volume of pores in substrate. Hence

$$C_{R0} = \frac{4N_{R0}}{\pi d^2 \delta_0 \phi_0} \tag{S4}$$

where $d$, $\phi_0$, and $\delta_0$ are the diameter, porosity, and thickness of the substrate respectively. Typically, the volume of reagent used for embedding $V_{reag}$ is same as the volume of pores in the substrate, to ensure uniform deposition. The analyte is provided through the droplet. If the concentration of analyte is $C_{A_{ppm}}$ in ppm, its concentration in mol/m³ is

$$C_{A0} = \frac{C_{A_{ppm}}}{M_{W_A}} \tag{S5}$$

where $M_{W_A}$ is the molecular weight of the analyte in g/mol.

*S.2.2 Analyte-Embedded (AE) Configuration*

This configuration is employed when the samples have very low concentrations of analyte. To preconcentrate the analyte, an adsorbent is added in the sample solution. Subsequently, the solution is filtered to get the residue which serves as the porous substrate. Let the concentration of the analyte in the sample solution be $C_{A_{ppm}}$ (in ppm) and the volume of sample is $V_{sample}$ in ml, then the number of moles embedded is

$$N_A = \frac{C_{A_{ppm}} \times V_{sample}}{M_{W_A}} \times 10^{-6} \tag{S6}$$

Here $M_{W_A}$ is the molecular weight of the analyte in g/mol. The concentration scale of the analyte $C_{A0}$ is defined as moles per unit volume of pores in the substrate. Hence

$$C_{A0} = \frac{4N_A}{\pi d^2 \delta_0 \phi_0} \tag{S7}$$

where $d$, $\phi_0$, and $\delta_0$ are the diameter, porosity, and thickness of the substrate respectively.

The reagent is provided through the droplet. Suppose the reagent concentration $C_{reag}$ is known in mg/ml, its concentration scale, $C_{R0}$ (in mol/m³) is

$$C_{R0} = \frac{C_{reag} \times 10^3}{M_{W_R}} \text{ mol/m}^3 \tag{S8}$$

where $Mw_R$ is the molecular weight of reagent in g/mol.

## S.3 Parameter Values

To determine characteristic concentration scales, the values of parameters required are given by **Table S1**. In this work, the model is applied in three contexts: (1) Theoretical analysis (Sections 3.2, 3.3 in Manuscript), (2) Experimental studies to validate in AE Configuration (Section 3.4.2 in Manuscript), and (3) Experimental studies to validate in RE Configuration (Section S.6 in SI). For the analysis and validation in RE configuration, the reaction between lead (analyte, $M_{w_A}$ = 207 g/mol) and sodium rhodizonate (reagent, $M_{w_R}$ = 214 g/mol) is employed. For validation in AE configuration, the Griess reaction for nitrite detection is employed.

**Table S1** Values of experimental parameters used in this study

| Parameter | Notation | Parameter value used for Analysis | Parameter value used for Validation (AE Configuration) | Parameter value used for Validation (RE Configuration) |
|---|---|---|---|---|
| Sample Volume | $V_{sample}$ | 25 ml (AE) | 30 ml | |
| Sample concentration | $C_{A_{ppm}}$ | 1 ppm (RE) <br> $10^{-3}$ ppm (AE) | 1, 100 ppm | 1, 10, 100 mM |
| Substrate diameter | $d$ | 5 mm | 14 mm | 4.2 mm |
| Substrate porosity | $\phi_0$ | 0.7 | 0.9 | 0.5 |
| Substrate thickness | $\delta_0$ | 180 μm | 180, 150, 110 μm | 180 μm |
| Initial droplet volume | $V_0$ | 2.5 μL | 17 μL | 5 μL |
| Initial droplet radius | $r_0$ | | 3.15 mm | 2.1 mm |
| Initial droplet contact angle | $\theta_{a0}$ | 20 degrees | | |
| Reagent stock concentration | $C_{reag}$ | $10^{-3}$ mg/ml (RE) <br> $10^{-2}$ mg/ml (AE) | 1 mg/ml, 8 mg/ml | 50 mM (Sulphanilamide) |

## S.3 Effect of concentration ratio $M_A$ on product distribution: Varying $C_{A0}$

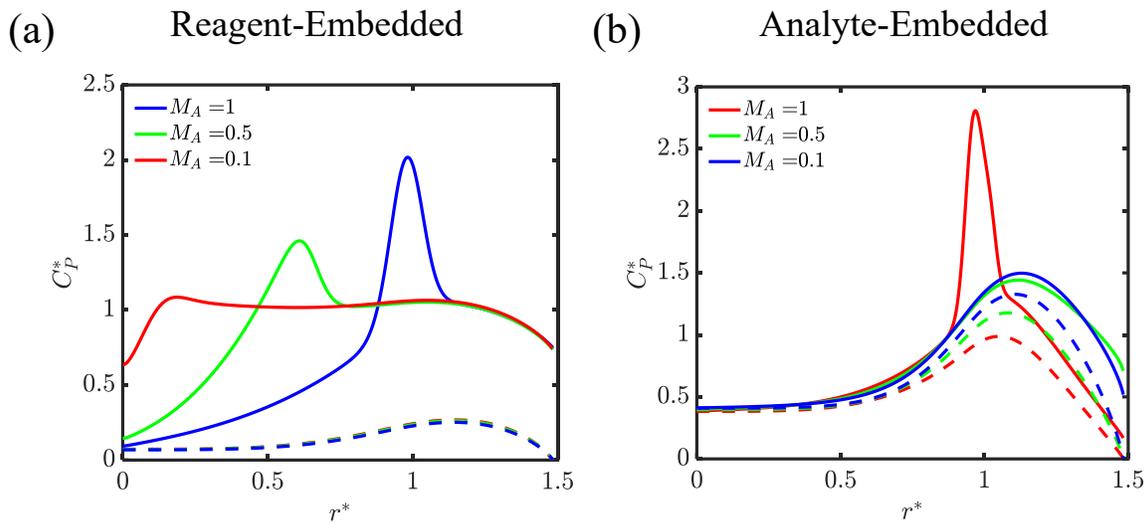

**Fig. S1** Effect of analyte-reagent concentration ratio ($M_A$) on product concentration for (a) RE, and (b) AE configurations. Here, $M_A$ is varied by varying $C_{A0}$. Dashed lines are at the end of Stage 1 and solid lines are at the end of stage 2.

## S.4 Effect of Immobilization: AE Configuration

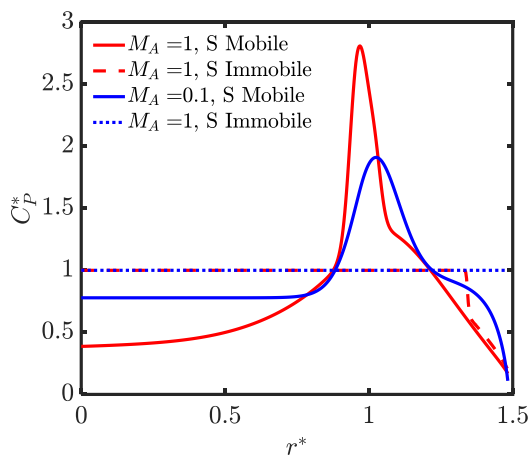

**Fig. S2** Effect of immobilization on the product distribution in AE configuration for different $M_A$. The product is immobile here.

## S.5 Materials and Methods

### S.5.1 Lead Detection

Lead ($Pb^{2+}$) in water samples is detected colorimetrically using sodium rhodizonate (NaRh) as a reagent. The reaction gives rise to a scarlet-colored precipitate, lead rhodizonate (PbRh). Stock solution of $Pb^{+2}$ (1000 mg $L^{-1}$) was prepared using lead nitrate ($Pb(NO_3)_2$) procured from Merck chemicals. Solutions of required concentrations in the range 1 ppm to 100 ppm were prepared by diluting with ultrapure de-ionized water (Milli-Q IQ7000, Millipore Corporation) before each experiment. To preconcentrate lead in a porous substrate, around 30 mg of kaolin adsorbent (LOBA Chemie (India)) was added to 30 mL of solution containing $Pb^{2+}$ at a fixed concentration. This was left aside for 15 minutes to allow the lead to be adsorbed on kaolin. The mixture was subsequently filtered using a Nylon filter membrane (Ultipor® N66 membrane disc filters from Pall® Corporation), and the resulting residue was used as the lead-enriched porous substrate for colorimetric analysis. The filter paper with the residue of the adsorbent layer was dried in an oven at 45 °C for 5 min. The dried sample was brought down to room temperature before adding the colorimetric reagent. The substrate formed has a diameter of 14 mm and a thickness of 110-180 µm. The reagent NaRh solution was prepared in a 1 ml buffer (pH 2.8) each time before the experiment. The buffer was made by adding 113 mg of L-tartaric acid (RANKEM™) and 58 mg of sodium hydrogen L-tartrate anhydrous (HiMedia) in 10 ml of distilled water. About 17 µL of the reagent drop was added on the substrate at the center, which reacts with $Pb^{+2}$ to form a scarlet-colored product. The process was monitored using a mobile phone from a height of 7 cm.

### S.5.2 Nitrite Detection

Nitrite ($NO_2^-$) detection was performed using the Griess reagent, which produces a pink colored azo dye upon reaction. The reagent consists of 50 mM sulphanilamide, 10mM n-(1-naphthyl) ethylenediamine dihydrochloride (NEDA) and 330 mM citric acid anhydrous in 80% methanol [1,2]. A 100 mM nitrite stock solution was prepared from sodium nitrite ($NaNO_2$) diluted with ultrapure deionized water (Milli-Q IQ7000) to obtain concentrations ranging from 1 to 100 mM. All the chemicals are procured from SRL.

A 5 µL droplet of the reagent was deposited onto a 1.5 × 1.5 cm² Whatman Grade 1 filter paper and allowed to dry at room temperature for 2 min. Subsequently, a 5 µL droplet of the analyte solution was added. Color development was recorded using a mobile phone camera positioned 7 cm above the substrate under uniform illumination.

### S.6 Experimental Measurements and Comparison with Theory (RE Configuration)

**Fig. S3** compares the experimentally observed color distributions with the product concentration profiles predicted by the model. The first column shows experimental images of the colored porous substrates. To quantitatively characterize the observed colorimetric response, RGB analysis was performed and a normalized relative color intensity $\theta$ is used to quantify the signal. This is defined as

$$\theta = \frac{255 - I_{green}}{255}$$

In the last column, radial distributions of product concentration predicted using the developed model are shown. The experiments were conducted for three nitrite concentrations (a) 1 mM, (b) 10 mM, (c) 100 mM to obtain three distinct characteristic distributions. In simulations, the parameter values (**Table S1**) used in the experiments were employed. For other parameters, values from the literature were adopted to qualitatively match predictions with experiments. The reaction follows second order kinetics i.e., first order with respect to nitrite and sulphanilamide [3,4]. The reaction rate constant $k_r$ was fixed at 0.1 mol$^{-1}$m$^3$s$^{-1}$. The mobility factors were set as $m_A = 1, m_R = 0.2$ and $m_A = 0$ for all three cases.

As shown in **Fig. S3**, the product distribution progressively shifts away from the center with increasing nitrite concentration (decreasing $M_A$). For case (a), corresponding to low nitrite concentration, the product concentration decreases monotonically with radial distance, indicating that the reaction predominantly occurs near the center. The model prediction closely matches with the measured $\theta$ profile. In case (b), a nearly flat color distribution is observed in the center upto $r/r_{max} \sim 0.7$, followed by a decline toward the periphery. The model qualitatively captures the trend. The onset of the decrease is predicted at a slightly smaller radius ($r/r_{max} \approx 0.5$). Case (c) represents a reagent-limited condition. Experimentally, a pale core followed by a pronounced ring is observed, with three regions in the $\theta$ profile. The flat core spreads till $r/r_{max} \sim 0.4$ and the peak maxima is located at

$r/r_{max} \sim 0.8$. The model predicts the distribution qualitatively with a wider ring. In the simulations the parameters used were from the literature and no attempt has been made to optimize the parameters to quantitatively match the predictions with experiments.

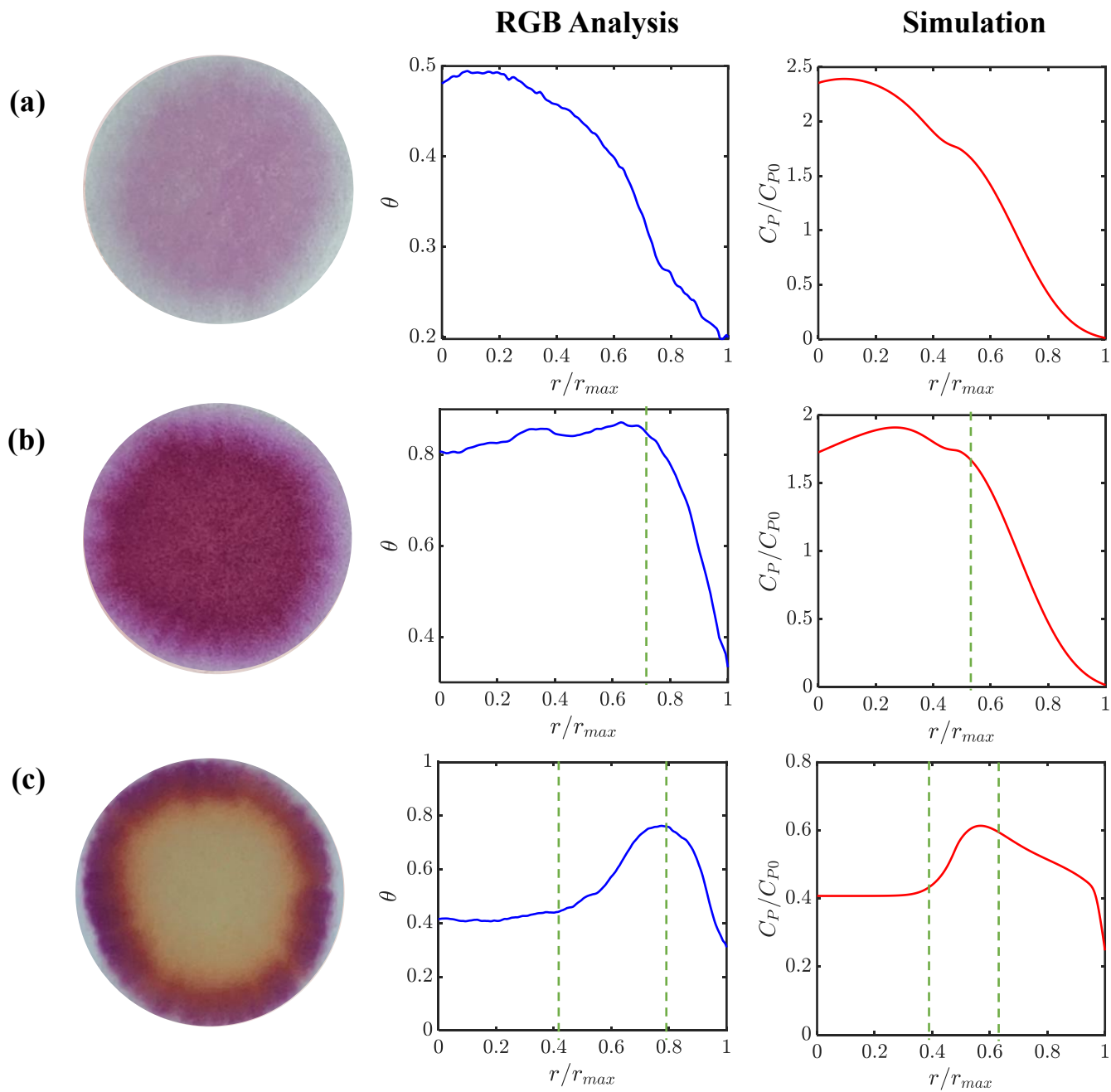

**Fig. S3** Comparison of spatial distributions of the color (product) observed experimentally in nitrite detection assay with model predictions for three different nitrite concentrations (a) 1 mM, (b) 10 mM, (c) 100 mM.